\documentclass[twocolumn,showpacs,showkeys,preprintnumbers,amsmath,amssymb,superscriptaddress,nofootinbib]{revtex4-2}
\usepackage{graphicx}
\usepackage{natbib}
\usepackage{amsmath}
\usepackage{epsfig}
\usepackage{amssymb}
\usepackage{xcolor}
\usepackage{cancel}
\usepackage{ulem}
\usepackage{lineno}
\usepackage{hyperref}

\begin{document}


\title{Isotopic Fission Yields of $^{240}$Pu as a Function of the Excitation Energy}

\author{D.~Ramos}
\email[]{diego.ramos@ganil.fr}
\affiliation{GANIL, CNRS/IN2P3, CEA/DRF, bd Henri Becquerel, 14076 Caen, France}

\author{M.~Caama\~no}
\affiliation{IGFAE - Universidade de Santiago de Compostela, E-15706 Santiago de Compostela, Spain}

\author{F.~Farget}
\affiliation{GANIL, CNRS/IN2P3, CEA/DRF, bd Henri Becquerel, 14076 Caen, France}

\author{C.~Rodr\'iguez-Tajes}
\affiliation{GANIL, CNRS/IN2P3, CEA/DRF, bd Henri Becquerel, 14076 Caen, France}

\author{A.~Lemasson}
\affiliation{GANIL, CNRS/IN2P3, CEA/DRF, bd Henri Becquerel, 14076 Caen, France}

\author{M.~Rejmund}
\affiliation{GANIL, CNRS/IN2P3, CEA/DRF, bd Henri Becquerel, 14076 Caen, France}

\author{C.~Schmitt}
\affiliation{IPHC Strasbourg, Universit\'e de Strasbourg CNRS/IN2P3, F-67037 Strasbourg Cedex 2, France}

\author{E.~Clement}
\affiliation{GANIL, CNRS/IN2P3, CEA/DRF, bd Henri Becquerel, 14076 Caen, France}

\author{L.~Audouin}
\affiliation{IJC Lab, Universit\'e Paris-Saclay, CNRS/IN2P3, F-91405 Orsay Cedex, France}

\author{J.~Benlliure}
\altaffiliation[Present address: ]{IFIC, Centro Mixto Universidad de Valencia-CSIC, Valencia, Spain}
\affiliation{IGFAE - Universidade de Santiago de Compostela, E-15706 Santiago de Compostela, Spain}

\author{E.~Casarejos}
\affiliation{CINTECX, Universidade de Vigo, DSN, Dpt. Mech. Engineering, E-36310 Vigo, Spain}

\author{D.~Cortina}
\altaffiliation[Present address: ]{IFIC, Centro Mixto Universidad de Valencia-CSIC, Valencia, Spain}
\affiliation{IGFAE - Universidade de Santiago de Compostela, E-15706 Santiago de Compostela, Spain}

\author{D.~Dor\'e}
\affiliation{CEA Saclay, DRF/IRFU/SPhN, 91191 Gif-sur-Yvette Cedex, France}

\author{B.~Fern\'andez-Dom\'inguez}
\affiliation{IGFAE - Universidade de Santiago de Compostela, E-15706 Santiago de Compostela, Spain}

\author{G.~de~France}
\affiliation{GANIL, CNRS/IN2P3, CEA/DRF, bd Henri Becquerel, 14076 Caen, France}

\author{A.~Heinz}
\affiliation{Chalmers University of Technology, SE-41296 G\"oteborg, Sweden}

\author{B.~Jacquot}
\affiliation{GANIL, CNRS/IN2P3, CEA/DRF, bd Henri Becquerel, 14076 Caen, France}

\author{C.~Paradela}
\altaffiliation[Present address: ]{EC-JRC, Institute for Reference Materials and Measurements, Retieseweg 1111, B-2440 Geel, Belgium}
\affiliation{IGFAE - Universidade de Santiago de Compostela, E-15706 Santiago de Compostela, Spain}

\author{T.~Roger}
\affiliation{GANIL, CNRS/IN2P3, CEA/DRF, bd Henri Becquerel, 14076 Caen, France}

\date{\today}

\begin{abstract}

Complete isotopic fission yields distributions of $^{240}$Pu have been measured as a function of the initial excitation energy. The $^{240}$Pu fissioning system was produced through a two-proton transfer reaction between a $^{238}$U beam and a $^{12}$C target. The reaction was measured in inverse kinematics at Coulomb barrier energies, allowing for the full distribution of fission fragments to be isotopically identified with the VAMOS++ Spectrometer. The excitation energy of the system was measured on an event-by-event basis by detecting the target-like recoil $^{10}$Be in a segmented silicon telescope. This manuscript reports on the evolution of the fission yields as a function of the excitation energy of the system between 8.2 to 11.9 MeV. The influence of the excitation energy is manifested in the damping of shell effects that feed the yields in the symmetry valley, as well as in a reduction of the neutron content of the fragments. This reduction, however, is observed only in the heavy fragment, while the neutron content of the light fragment remains unaffected. The comparison with previous measurements, models, and evaluations highlights the importance of correlated observables for improving fission models. 

\end{abstract}

\keywords{}

\pacs {}

\maketitle

\section{INTRODUCTION}


The fission process is a good laboratory for studying dynamical effects of the nuclear matter and nuclear structure. The non-equilibrium collective motion between the saddle and scission deformations in fission is influenced by dissipation and inertia, which lead to fluctuations and correlations~\cite{SCHUNCK2022103963,10.3389/fphy.2020.00063}. On the other hand, nuclear structure favors specific configurations of the potential energy landscape followed by the system, which influences the relative production of fission fragments~\cite{wilkins,SPY,schGEF}. Fission yields of actinides are a clear proof of the effect of nuclear structure on the fission process: at low excitation energies, nuclear shells cause predominatly asymmetric fission.

However, the strong interplay between these two aspects of the fission process, together with its intrinsic complexity, has so far prevented a fully microscopic description. Only recently, the most advanced microscopic approaches are able to describe the evolution of the fissioning system from the saddle point to the scission point~\cite{Bulgac,Tanimura}, and the impact of specific nuclear configurations with a given number of protons, neutrons and defined deformations was found to be responsible for asymmetric fission~\cite{Scamps,ScampsPRC}. Still, questions regarding the fission process remain open. For instance, the moment along the fission path when pre-fragments exhibit a well-defined individual structure is still unclear. The dissipative character of the process is also under debate~\cite{ram23} as well as the generation of angular momentum~\cite{WilsonNature} and the competition with other decay channels throughout the process~\cite{MultiChanceFission,Abdurrahman24}. 

From the experimental point of view, precise measurements of fission-fragment observables help to guide and challenge models. The historical limitation of access to only long-lived fissioning systems and fragment masses was lifted, on the one hand, by the surrogate technique, which provided access to exotic systems~\cite{Esch12,Nishio16}, and on the other hand, by the inverse-kinematics technique, which allowed for the measurement of the proton numbers of the full distribution of fission fragments~\cite{Sch00,Morfouace2025}.  
 
The use of magnetic spectrometers in inverse kinematics has significantly increased the number of fission-fragments observables allowing for the simultaneous measurement of the proton and mass numbers of the full distribution of fission fragments~\cite{caaPRC13,pell17,ramPRC18,chat20,SOFIA21}.

Heavy-ion beams at Coulomb barrier energies have additionally provided access to the excitation energy of the system by using multi-nucleon transfer reactions in inverse kinematics to populate fissioning systems~\cite{rodPRC14}. This method has been recently implemented in storage rings with promising results~\cite{JuradoStoring}. 

In this context, a fission program based on multi-nucleon transfer-induced fission and fusion-induced fission has been running at GANIL for more than a decade. Stable heavy beams at Coulomb barrier energies are used to induce fission in inverse kinematics. The full distribution of fission fragments from exotic fissioning systems is isotopically identified using the VAMOS++ magnetic spectrometer~\cite{rejNIM}, together with a precise determination of the initial excitation energy of the system. This experimental program has provided access to one of the largest sets of correlated experimental data related to fission fragments~\cite{caaPRC13,Caa15,rodPRC14,CAAPLB,ramPRC18,ramPRC19,ramPRL,ramPRC20,Sch21,SchmittPRC, CoboProceeding}, including pre- and post-neutron evaporation fission yields, prompt-neutron multiplicity, total kinetic and excitation energies of fission fragments, proton even-odd staggering, and scission configurations. 

A further step in this program consists in a detailed study of the isotopic population of fission fragments as a function of the system excitation energy, providing simultaneous insight into the damping of shell effects and the evolution of the neutron content of the fragments.

\begin{figure}[!]
\includegraphics[width=0.49\textwidth]{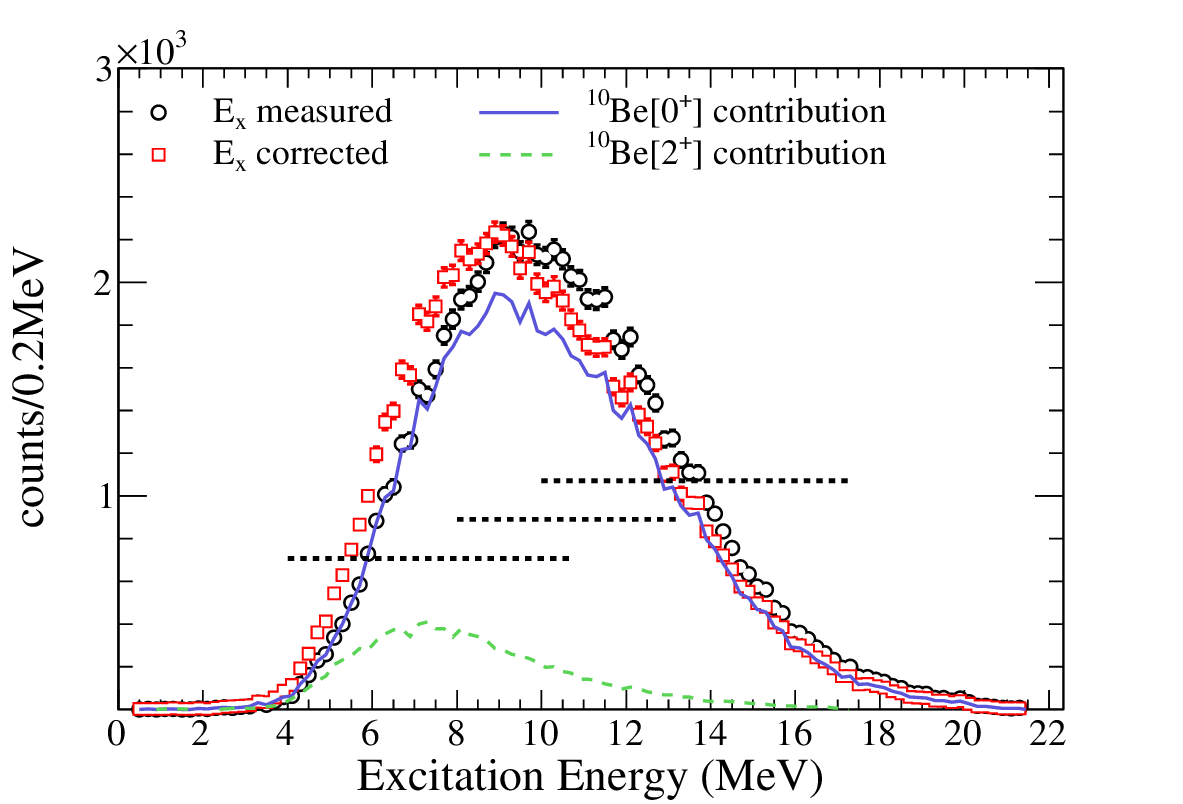}
\caption{Excitation energy distribution of $^{240}$Pu obtained in the present work, measured in coincidence with the detection of fission fragments (black circles)
and corrected for the probability of the population of excited states in $^{10}$Be (red squares). The contribution of the ground state ($0^+$) and the first excited state ($2^+$) of $^{10}$Be are indicated with solid blue and dashed green lines, respectively. Dotted black lines represent the different excitation energy ranges selected for this study.}
\label{fig:Ex}
\end{figure}

This manuscript reports on the isotopic fission yields of $^{240}$Pu, populated through 2-proton-transfer-induced fission between a $^{238}$U beam and a $^{12}$C target, measured at the focal plane of the VAMOS++ spectrometer. Three weighted-average excitation energies are investigated in this work: 8.2, 10.0, and 11.9~MeV. Mass and elemental fission yields are presented and compared with existing data from neutron-induced fission and with GEF calculations~\cite{schGEF}, a semi-empirical scission-point model that accurately reproduces the main features of fission fragments. The neutron content of post-neutron-evaporation fission fragments is determined thanks to the simultaneous measurement of mass and proton numbers of the full fragment distributions. The total prompt-neutron multiplicity distribution is also presented and discussed as a function of the fragment atomic number and excitation energy.

\section{EXPERIMENTAL METHOD AND ANALYSIS}

\begin{figure*}[!]
\includegraphics[width=1\textwidth]{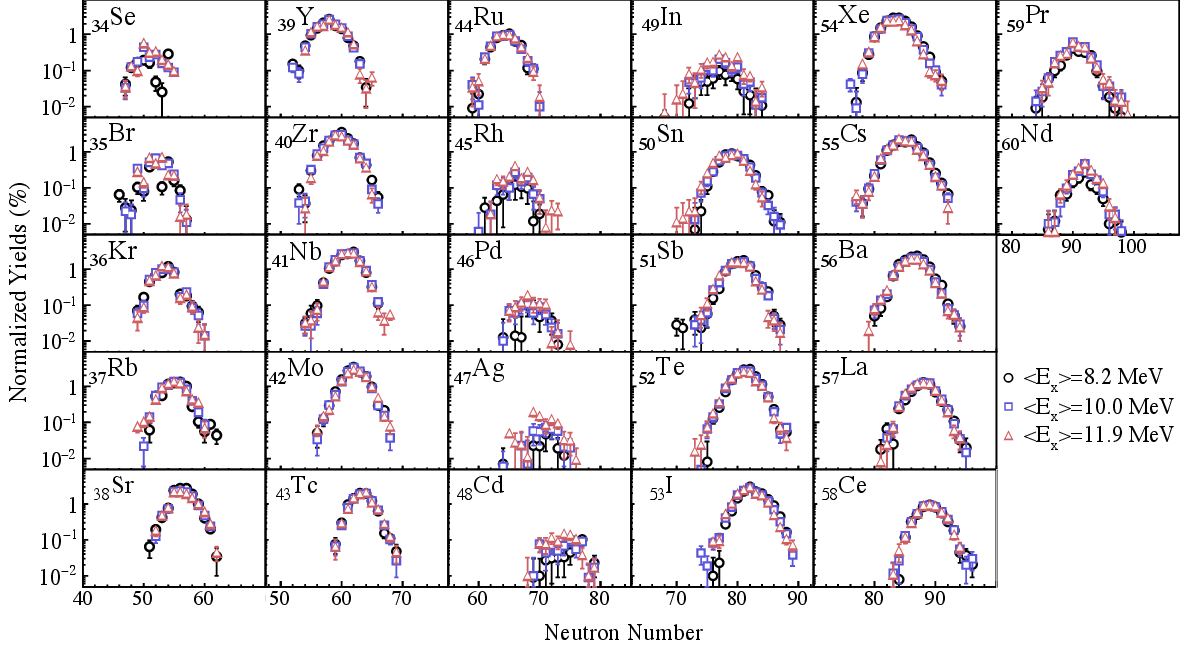}
\caption{Normalized isotopic fission yields of $^{240}$Pu. The three $E_x$ data sets ---8.2~MeV, 10.0~MeV, and 11.9~MeV--- are represented with black cicles, blue squares, and red triangles, respectively. Each pannel corresponds to one element which is indicated in the upper-left side of the pannel. Yields are presented as a function of the neutron number of the fission fragments.}
\label{fig:ISO}
\end{figure*}

\begin{table} [!t]
\begin{center}
  \caption{\label{table1}Characteristics of the excitation energy ranges of $^{240}$Pu selected for the analysis are presented. 
These ranges encompass the limits of each range, along with the weighted-average values before and after correction for the target-like recoil excitation. 
Additionally, standard deviations are provided for each range.}
  \begin{tabular}{|c|c|c|c|}
    \hline\hline
    $E_{x}$ limits  &~$\langle E_{x}^m\rangle$ & $\langle E_{x}\rangle$ &  $SD_{E_{x}}$\\
   (MeV) & (MeV) & (MeV) & (MeV) \\\hline
4.0 - 10.7 & 8.5 & 8.2 & 1.46\\
8.0 - 13.3 & 10.5 & 10.0 & 1.46\\
10.0 - 17.3 & 12.5 & 11.9 & 1.79\\\hline

  \end{tabular}
\end{center}
\end{table}

The present measurement was conducted at GANIL, France, where a beam of $^{238}$U was accelerated to $6.14$~AMeV and 
impinged on a $100~\mu$g/cm$^2$-thick $^{12}$C target. The actinide $^{240}$Pu was produced through two-proton-transfer reactions
$^{12}\text{C}(^{238}\text{U},^{240}\text{Pu}^*)^{10}\text{Be}$ in inverse kinematics, and underwent fission in flight. 
The fissioning system 
was identified by detecting the target-like recoil $^{10}$Be, using the SPIDER silicon telescope, as described in Ref.~\cite{rodPRC14}. Its excitation 
energy, $E_x$, was determined on an event-by-event basis by reconstructing the binary reaction using energy and momentum conservation laws, as follows: 
\begin{equation}
\label{eq_ex}
\begin{aligned}
& E_x = Q_{gg}+E_{beam}-E_{^{10}Be}-M_{^{240}Pu}^2+p_{^{240}Pu}^{2}+M_{^{240}Pu},\\
& p_{^{240}Pu}^{2} =  p_{beam}^{2} +  p_{^{10}Be}^{2} - 2 p_{beam} p_{^{10}Be} \cos(\theta_{^{10}Be});
\end{aligned}
\end{equation}
with $Q_{gg}$ being the ground-state to ground-state reaction Q value; $E_{beam}$ and $p_{beam}$ the kinetic energy and momemtum of the beam; $E_{^{10}Be}$ and  $p_{^{10}Be}$ the kinetic energy and momentum of of the target-like recoil $^{10}Be$; and $\theta_{^{10}Be}$ the angle of $^{10}Be$ with respect to the beam direction. The resulting $E_x$ distribution ranges from $4$ to $20$~MeV with a resolution of $\sigma=1.15$~MeV.

The distribution of $E_x$ of $^{240}$Pu measured in this 
experiment in coincidence with fission events, $f^{m}_f(E_x)$, is shown with black circles in Fig.~\ref{fig:Ex}. This distribution includes a contributon from the first excited state ($2^{+}$) of $^{10}$Be, with a measured probability of $0.14 \pm 0.04$~\cite{rodPRC14}, obtained using the HPGe EXOGAM detector~\cite{EXOGAM}. The excitation of the target-like recoil results in an effective reduction of the excitation energy available for $^{240}$Pu. To correct for this effect, the measured distribution is fitted with the sum of the 
ground state ($f^{0^+}_f(E_x)$) and the first excited state ($f^{2^+}_f(E_x)$) contributions, the latter shifted by 3.37 MeV, corresponding to the energy of the first excited state of $^{10}$Be:
\begin{equation}
\label{eq_ex1}
f^{m}_f(E_x) = f^{0^+}_f(E_x)+f^{2^+}_f(E_x+3.37 \text{ MeV}),
\end{equation}
where
\begin {equation}
\label{eq_ex2}
\begin{aligned}
& f^{0^+}_f(E_x) = 0.86P_f(E_x)f^{m}_{tot}(E_x), \\
&f^{2^+}_f(E_x)=0.14P_f(E_x-3.37 \text{ MeV})f^{m}_{tot}(E_x-3.37 \text{ MeV}), 
\end{aligned}
\end {equation}
with $P_f(E_x)$ being the fission probability and $f^{m}_{tot}(E_x)$ the total $E_x$ distribution. 

The corrected $E_x$ distribution, $f^{c}_f(E_x) =  f^{0^+}_f(E_x)+f^{2^+}_f(E_x)$, is presented with red squares in Fig.~\ref{fig:Ex}. The blue solid and the green dashed lines represent the contributions of the ground state and the first excited state of $^{10}$Be to the corrected distribution.

The selection of $^{10}$Be events includes $5\pm 2$~\% of contamination due to overlap with $^{9}$Be~\cite{rodPRC14}. The subtraction of this contamination results in a maximum shift of 0.04~MeV towards lower $E_x$. This value is two orders of magnitude smaller than the experimental resolution and it is therefore neglected in the calculation of $E_x$.

The ranges of $E_x$ selected for this study were determined under the condition of having $4\times10^4$ events per $2$-MeV step in the weighted-average $E_x$. They are shown with black dotted lines in Fig.~\ref{fig:Ex} and listed in Table~\ref{table1}.

When decaying in flight, the fissioning system splits into two fragments emitted within a cone of $30$ degrees in the laboratory frame.
For each fission event, one of the fission fragments within the detection acceptance was detected at the focal-plane setup of the magnetic spectrometer VAMOS++ between 150~ns and 300~ns after the fission reaction, and it was isotopically identified~\cite{rejNIM,LemNIM}. The spectrometer acceptance and efficiency were determined using a self-consistent method, as described in~\cite{ramPRC18}.

 The isotopic fission yields of 
$^{240}$Pu ($Y(Z,A,E_x)$), normalized to 200\%, were independently calculated for each range of $E_x$ as follows:
\begin{equation}
\label{eq_Y}
 Y(Z,A,E_x) = 200\frac{N(Z,A,E_x)}{\sum_{Z,A}{N(Z,A,E_x)}},
\end{equation}
where $N(Z,A,E_x)$ is the number of events of each isotope $(Z,A)$ detected in the focal plane of the spectrometer and corrected for acceptance and efficiency, for a given excitation energy $E_x$.
The uncertainties of the fission yields were determined as the quadratic sum of statistical and systematic contributions. Systematic uncertainties range from 4\% to 10\% for the largest yields, taking into account normalization, intrinsic efficiency, and spectrometer acceptance uncertainties. 

The effect of contamination from $^{241}$Pu due to the overlap between $^{10}$Be and $^{9}$Be in the isotopic fission yields was evaluated using the GEF code. This contamination causes the fission yields to vary from 0.5\% to 1.5\%. This error was added to the uncertainty representing less than 10\% of the total uncertainty.


\begin{figure*}[!]
\includegraphics[width=1\textwidth]{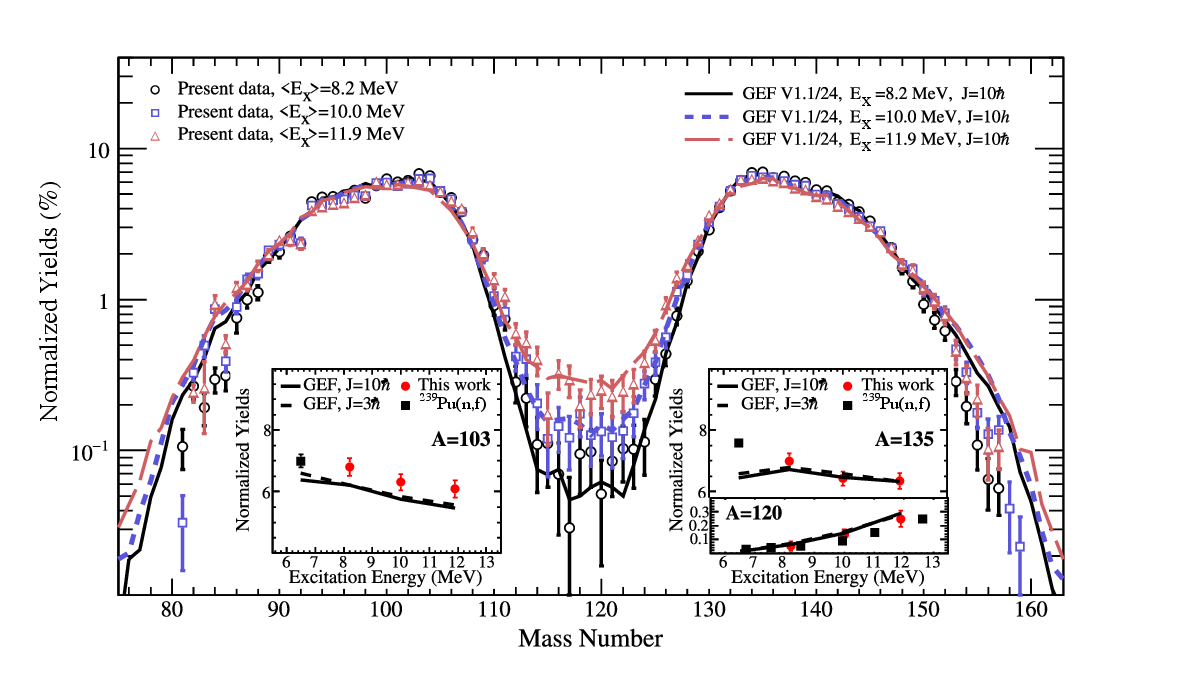}
\caption{Fission yields of $^{240}$Pu as a function of the fragment masses. Data at three excitation energies, 8.2, 10.0, and 11.9~MeV are presented with black circles, blue squares, and red triangles, respectively. GEF calculations at the respective three excitation energies and average initial spin $J=10\hbar$ are also presented with black solid, blue dotted, and red dashed lines. Insets: Evolution of the fission yields of the most populated masses A=103 and A=135, and the mass at the symmetry A=120 as a function of $E_x$. Present data (red dots) are compared with data from neutron-induced fission from Refs.~\cite{schnpa,BaiPRC,GinPRC} (black squares) and with GEF predictions with initial spin $J=10\hbar$ (solid line) and $J=3\hbar$ (dotted line).}
\label{fig:YA}
\end{figure*}

\section{ISOTOPIC FISSION YIELDS}

Independent isotopic fission yields of $^{240}$Pu are presented in Figure~\ref{fig:ISO}~\cite{Supplementary}. Elements from Se to Nd are shown in separated panels as a function of their neutron number. The three $E_x$ values, 8.2, 10.0, and 11.9~MeV are depicted with black circles, blue squares, and red triangles, respectively. The sensitivity of this measurement allows to measure fission yields as small as $10^{-2}$\%, corresponding to cross sections smaller than 1$\mu$b. Fluctuations are observed in the lighter elements, namely Se and Br, with a contribution to the elemental and mass fission yields of less than 1\%. Asymmetric fission is clearly visible, showing reduced fission yields at elements around Ag and higher fission yields around Zr and Xe. 

When the excitation energy increases, the elements located in the valley, namely Rh, Pd, Ag, Cd, and In, show an enhancement of their respective yields. This reflects a clear damping of shell effects. Another effect of $E_x$ is observed in the heavy-fragment region, with a shift toward less neutron-rich isotopes as $E_x$  increases. In contrast, this effect is not observed in the light-fragment region. This indicates an increase in neutron evaporation from the heavy fragments, while the light fragments remain unaffected, as will be discussed.


\section{MASS AND ELEMENTAl FISSION YIELDS}

Post-neutron-evaporation mass fission yields ($Y(A,E_x)$) of $^{240}$Pu are calculated as the sum of isotopic fission yields over the different elements of each isobar, $Y(A,E_x) = \sum_{Z}{Y(Z,A,E_x)}$. These mass fission yields are presented in Fig.~\ref{fig:YA}. The three data sets at 8.2, 10.0, and 11.9 MeV of excitation energy are shown with black circles, blue squares, and red triangles, respectively. 

Present data are compared with calculations from the GEF code (V1.1/2024) at the respective $E_x$. Previous measurements using a similar two-proton-transfer induced fission~\cite{Tanaka22} suggest an upper limit of $J=10\hbar$ for the initial spin of the system, compared to $J=3\hbar$, expected for neutron-induced fission at these energies. The GEF calculations are therefore performed with $J=10\hbar$. As discussed later, GEF predicts negligible differences with respect to $J=3\hbar$ within the range of $E_x$ investigated in this work. These results are presented in black solid, blue dotted, and red dashed lines, corresponding to increasing $E_x$, respectively.

Both the present data and the GEF code show similar feeding of the symmetry valley with increasing $E_x$. This suggests a good modeling of the damping of shell effects by GEF. However, some discrepancies are still observed between present data and the model, particularly at very asymmetric splits where the model overestimates the yields.

\begin{figure}[!]
\includegraphics[width=0.49\textwidth]{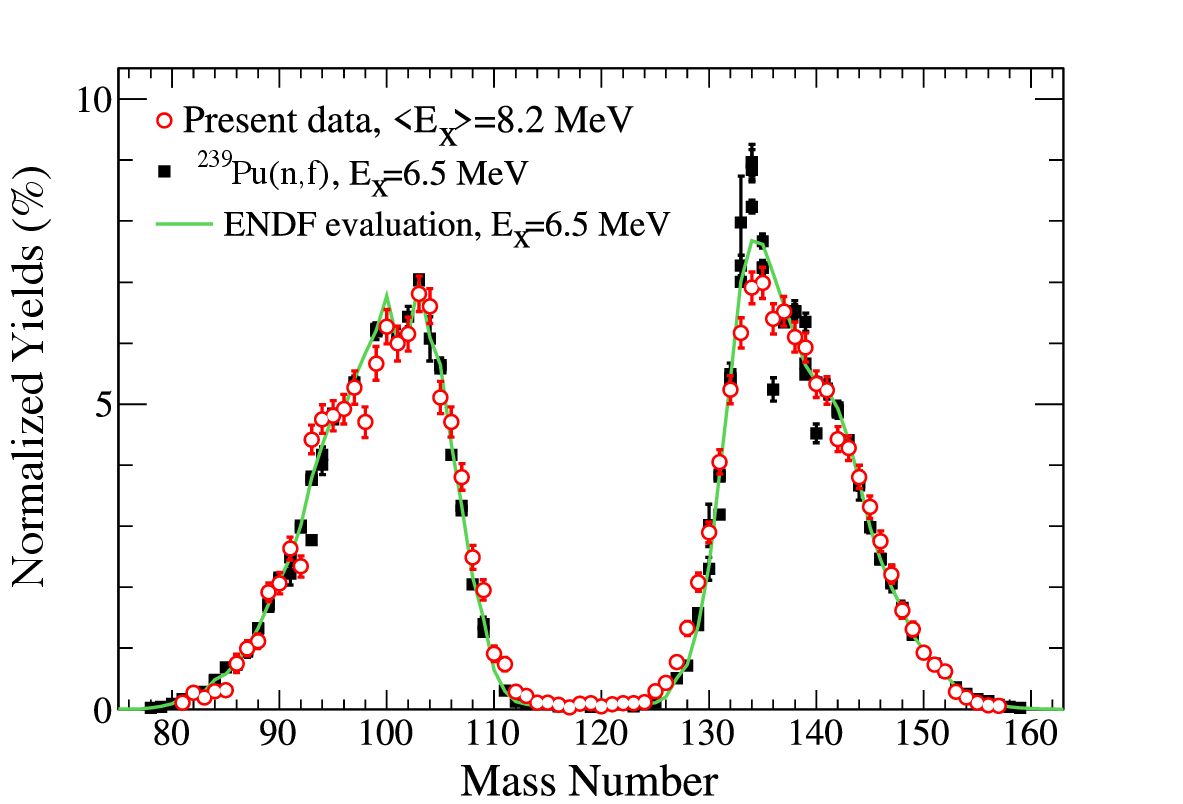}
\caption{Fission yields of $^{240}$Pu as a function of the fragment mass numbers at $E_x=8.2$~MeV, reported in this work (red cicles) compared with that from thermal-neutron-induced fission of $^{240}$Pu from Ref.~\cite{Naik03072022} (black squares) and to fission yields evaluation of ENDF/B-VI from Ref.~\cite{ENDF_PU} (green line).}
\label{fig:YA_compare}
\end{figure}

\begin{figure}[!]
\includegraphics[width=0.49\textwidth]{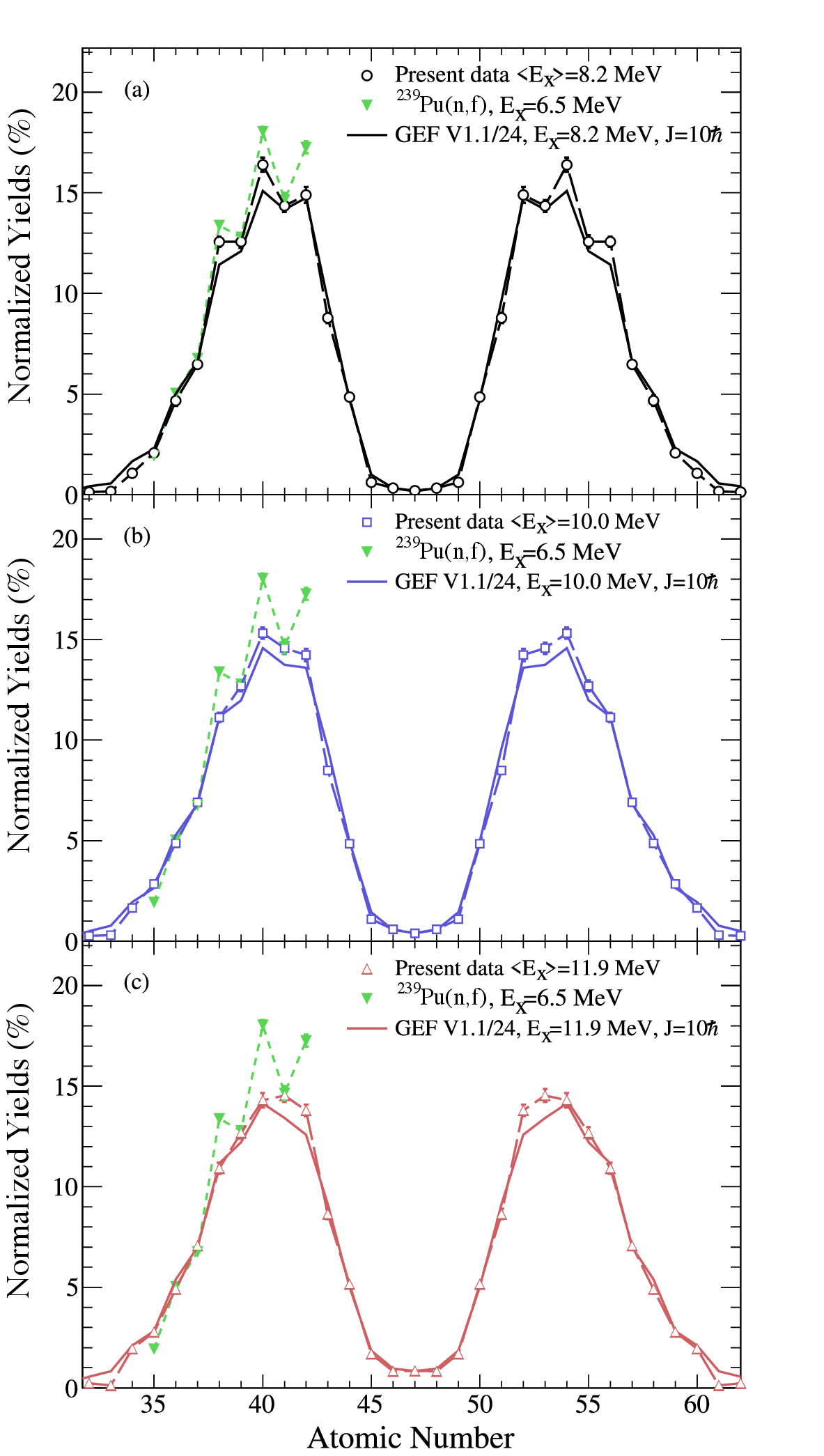}
\caption{Elemantal fission yields of $^{240}$Pu at three excitation energies: 8.2 MeV (a), 10.0 MeV (b), and 11.9 MeV (c). GEF calculations at these excitation energies and initial spin $J=10\hbar$ are also presented with solid lines. Data from thermal-neutron-induced fission of $^{239}$Pu from Ref.~\cite{schnpa} are also included (green inverted triangles). Dashed lines connect the experimental points to guide the eye.}
\label{fig:YZ}
\end{figure}

The insets of Fig.~\ref{fig:YA} show the evolution of the fission yields as a function of $E_x$ for the most populated light and heavy masses, $A=103$ and $A=135$, as well as for the mass at the symmetry, $A=120$. The data from this work (red dots) are compared with data from neutron-induced fission from Refs.~\cite{schnpa,BaiPRC,GinPRC} (black squares) and with two GEF calculations: one with initial spin $J=10\hbar$ (solid line) and another with initial spin $J=3\hbar$ (dashed line). At masses $A=103$ and $A=135$, both sets of data exhibit a monotonous trend: the higher the $E_x$, the lower the fission yields. In contrast, GEF shows a sudden yield reduction at the lower $E_x$, more pronounced for $J=10\hbar$. Above $E_x=6.5$~MeV, a good agreement is achieved between the data and the model in the heavy mass, $A=135$. In the light mass, $A=103$, the model predicts systematically lower yields than those observed in the data. At symmetry, present data are in good agreement with both previous neutron-induced fission data and GEF predictions, which turn out to be relatively insensitive to $J$ in the studied range of $E_x$.      

\begin{figure*}[!]
\includegraphics[width=0.7\textwidth]{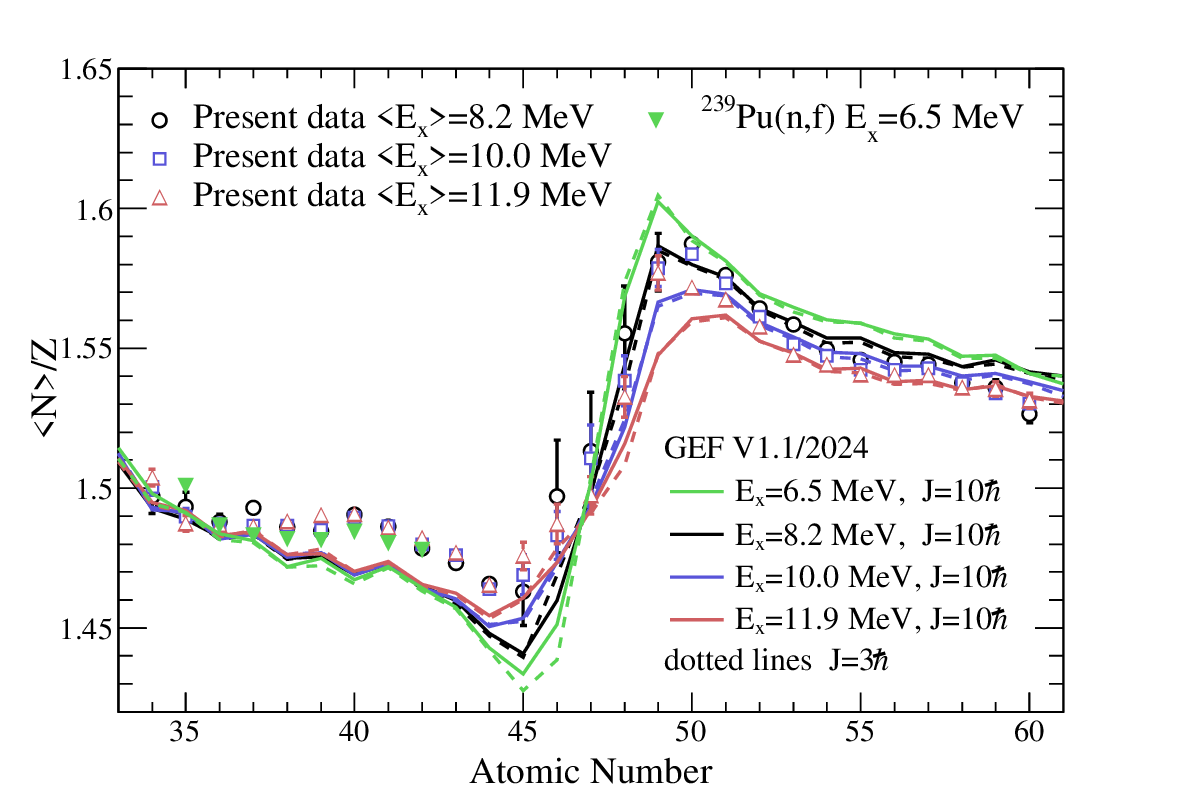}
\caption{Post-neutron-evaporation neutron excess of fission fragments as a function of their atomic number. Present data at three $E_x$, 8.2, 10.0,and 11.9~MeV are presented with black circles, blue squares, and red triangles, respectively. Data from thermal-neutron-induced fission, leading to 6.5~MeV of $E_x$, from Ref.~\cite{schnpa} are included and represented with green inverted triangles. GEF calculations at the same $E_x$ are indicated with solid lines for a initial average spin population of $J=10\hbar$ and with dotted lines for a spin of $J=3\hbar$ following the same color pattern as the experimental data.}
\label{fig:NZ}
\end{figure*}

The data set of this work at the lower $E_x$ ($8.2$~MeV) is compared with previous measurement of mass yields from thermal-neutron-induced fission of $^{239}$Pu from Ref.~\cite{Naik03072022} where the fission yields were deduced from cumulative yields ---i.e. the fission-fragment population after $\beta$-decay---. The comparison is presented in Fig.~\ref{fig:YA_compare} where present data are depicted with red circles and previous measurement with black squares. An overall good agreement between both sets of data is observed with a main discrepancy at A$\approx$134. The enhancement of yields observed in neutron-induced fission data around A$\approx$134 is a characteristic of cumulative yields as the end point of the $\beta$-delayed neutron emission of heavier neutron-rich isotopes, suggesting that part of the shift due to the $\beta$-delayed neutron emission was not fully corrected when computing post-neutron-evaporation yields. The discontinuity observed at $A=92-93$ is present in both sets of data although more pronounced in the present data. The effect of $E_x$ is evidenced in the descent from the maximum yields towards the symmetry valley where thermal-neutron-induced fission data ($E_x=6.5$~MeV) show a steeper slope than that in the present data at $E_x=8.2$~MeV.

The ENDF/B-VI evaluation of mass fission yields for thermal-neutron-induced fission of $^{239}$Pu, from Ref.~\cite{ENDF_PU}, is included in Fig.~\ref{fig:YA_compare} and is shown with a green line. Overall, good agreement is observed between present data and the evaluation although some differences exist, such as the higher yields at A=134, which can be attributed to differences in prompt-neutron evaporation resulting from the different excitation energies.    

Elemental fission yields are calculated in a manner analogous to mass fission yields, as the sum of the isotopic fission yields over the different isotopes of a given element, $Y(Z,E_x) = \sum_{A}{Y(Z,A,E_x)}$. They are presented in Fig.~\ref{fig:YZ} for the three $E_x$ studied in this work: 8.2 MeV (a), 10.0 MeV (b), and 11.9 MeV (c). Data from thermal-neutron-induced fission of $^{239}$Pu from Ref.~\cite{schnpa}, available only for a limited range of light fragments, are also included and depicted with green triangles. 

The effect of $E_x$ is evident in the symmetric fission region which, as expected, follows the same trend as the mass yields, with the valley being increasingly filled at higher $E_x$.

Elemental fission yields show a maximum for the element Xe ($Z = 54$) at low excitation energy which shifts towards the element I ($Z = 53$) as the excitation energy increases. This behavior is driven by the reduction of the pronounced even-odd staggering, with yields of even-protons nuclei being systematically enhanced compared to those of odd-proton nuclei. This effect is a signature of the intrinsic excitation energy available from the saddle to the scission point. This topic has been addressed in a previous publication~\cite{ram23} and will not be discussed further here.

The present data are compared with GEF calculations at the same $E_x$ and initial spin $J=10\hbar$. Excellent agreement is observed in the symmetry valley, with a similar damping of shell effects. GEF also reproduces well the reduction of the proton even-odd staggering, with some differences in the production, mainly between $Z=52$ and $Z=54$.

\section{FISSION-FRAGMENT NEUTRON EXCESS, NEUTRON EVAPORATION, AND ISOTONIC FISSION YIELDS}

The neutron excess of fission fragments ($\langle N\rangle/Z$) is defined as the ratio of the average number of neutrons to the number of the protons of each fission-fragment element. This is calculated in this work from post-neutron-evaporation isotopic fission yields as~\footnote{Each $\langle N\rangle/Z$ value is recalculated $10^6$ times by randomly varying the yield values within their uncertainties. The standard deviation of the resulting $\langle N\rangle/Z$ distribution is adopted as the final $\langle N\rangle/Z$ uncertainty.}: 
\begin{equation}
\frac{\langle N\rangle}{Z}(Z,E_x)= \frac{1}{Z} \frac{\sum_A{[A-Z]\times Y(Z,A,E_x)}}{\sum_A{Y(Z,A,E_x)}}.
\end{equation}

The present data are shown in Fig.~\ref{fig:NZ} with black circles, blue squares, and red triangles, corresponding to data sets at $E_x=8.2$, 10.0, and 11.9~MeV, respectively. Data show a pronounced charge polarization. Heavy fragments are systematically more neutron rich than light fragments. The most neutron-rich isotopes are produced around $Z\approx 50$, consistent with previous measurements~\cite{caaPRC13,pell17}. This result highlights the importance of correlated observables to disentangle the impact of structural effects: while the maximum of fission yields appears at $Z\approx 54$, the largest deviation from liquid drop-behavior in $N/Z$ appears at $Z\approx 50$, presumably driven by the doubly magic nucleus $^{132}$Sn.

The impact of increasing initial $E_x$ is reflected in the heavy fragment with a continuous reduction of the neutron-richness of the fragments. In contrast, the neutron excess of light fragments exhibit stability against variations in $E_x$. This observation suggests that the neutron excess at scission does not change significantly within the excitation energy range studied in this work, while the additional initial $E_x$ is dissipated through neutron evaporation primarily from the heavy fragment. This is consistent with theoretical considerations suggesting that the intrinsic excitation energy introduced in the system in a superfluid regime is stored in the heavy pre-fragments before scission~\cite{EnergySorting}. 

Data from thermal-neutron-induced fission reported in Ref.~\cite{schnpa}, limited to the light fragment, are also presented in Fig.~\ref{fig:NZ} with green triangles. When comparing with the present data, both show a local maximum at $Z=40$. However, present data are slightly more neutron rich overall compared to the thermal-neutron-induced fission results. This effect is attributed to a reduced neutron evaporation from the saddle to the scission point as well as from the fission fragments due to the additional angular momentum present in the system induced by the two-proton-transfer reaction compared to neutron-capture reactions, as reported in Ref.~\cite{RamosPRL25}.  

Figure~\ref{fig:NZ} also includes calculations from GEF at the same $E_x$ of the experimental data, namely 8.2 MeV (black line), 10.0 MeV (blue line), and 11.9 MeV (red line), as well as 6.5 MeV (green line). Two calculations with initial spins $J=10\hbar$ (solid lines) and $J=3\hbar$ (dotted lines) were computed. When comparing the calculation with experimental data, a better agreement is observed in the heavy fragments than in the light ones. The largest differences between present data and the model are observed in the light fragments around $Z\approx 40$ where the model predicts systematically less neutron-rich fragments compared to data. The same behaviour is observed with thermal-neutron induced fission. 

The effect of the initial excitation energy, however, is generally well reproduced by the model, showing a negligible effect in the light fragment region and a continuous decrease of the neutron content in the heavy fragment. In the region around $Z\approx 50$ the model predicts less neutron rich isotopes compared to data.  

According to GEF, the effect of the initial spin is an increase in the neutron content of the fragments, mainly around symmetry, with a minor impact. 
The large uncertainties of the data prevent from an accurate comparison at the symmetry.

\begin{figure}[!]
\includegraphics[width=0.49\textwidth]{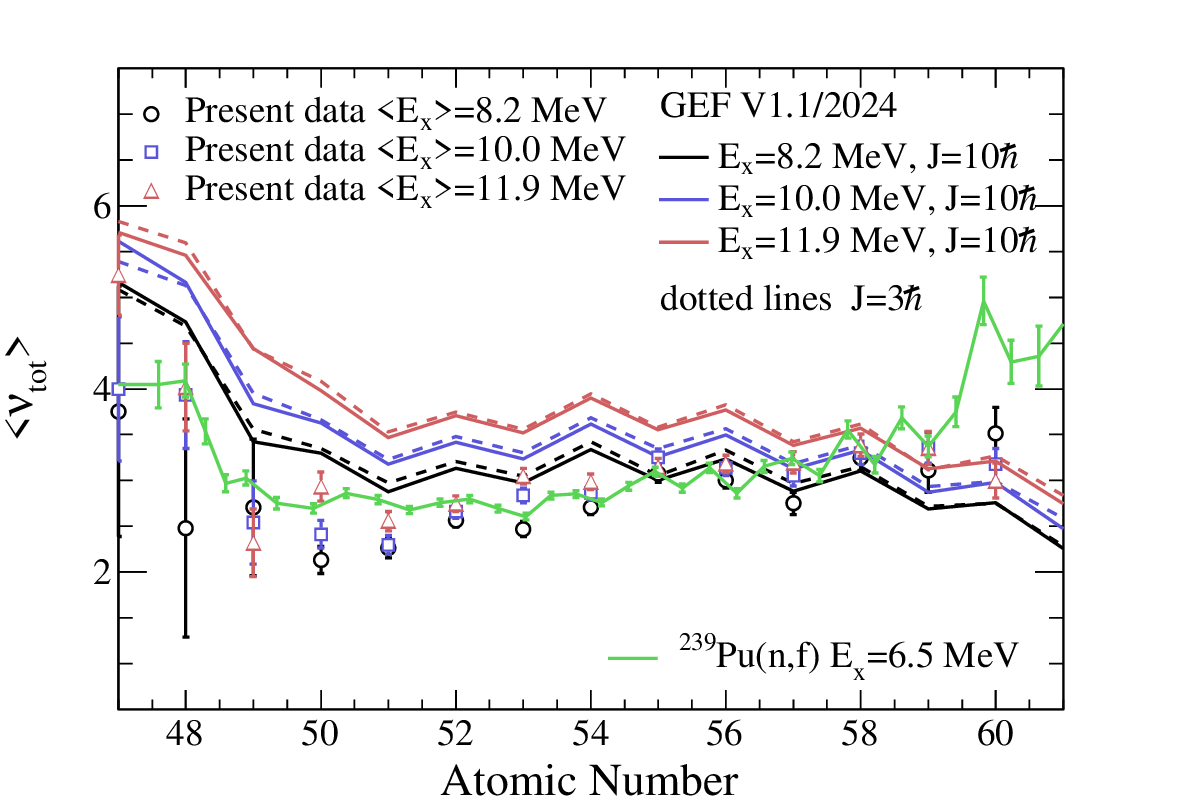}
\caption{Total prompt neutron multiplicity of $^{240}$Pu as a function of the atomic number of the heavy fragments. Present data at three $E_x$, 8.2, 10.0,and 11.9~MeV are presented with black circles, blue squares, and red triangles, respectively. GEF calculations at the same $E_x$ are indicated with solid lines for a initial spin of $J=10\hbar$ and with dotted lines for $J=3\hbar$ following the same color pattern as the experimental data. Adapted data of thermal-neutron-induced fission from Ref.~\cite{TSUCHIYA01112000} is also included and represented with green solid line.}
\label{fig:Nevap}
\end{figure}

\begin{figure*}[!]
\includegraphics[width=0.9\textwidth]{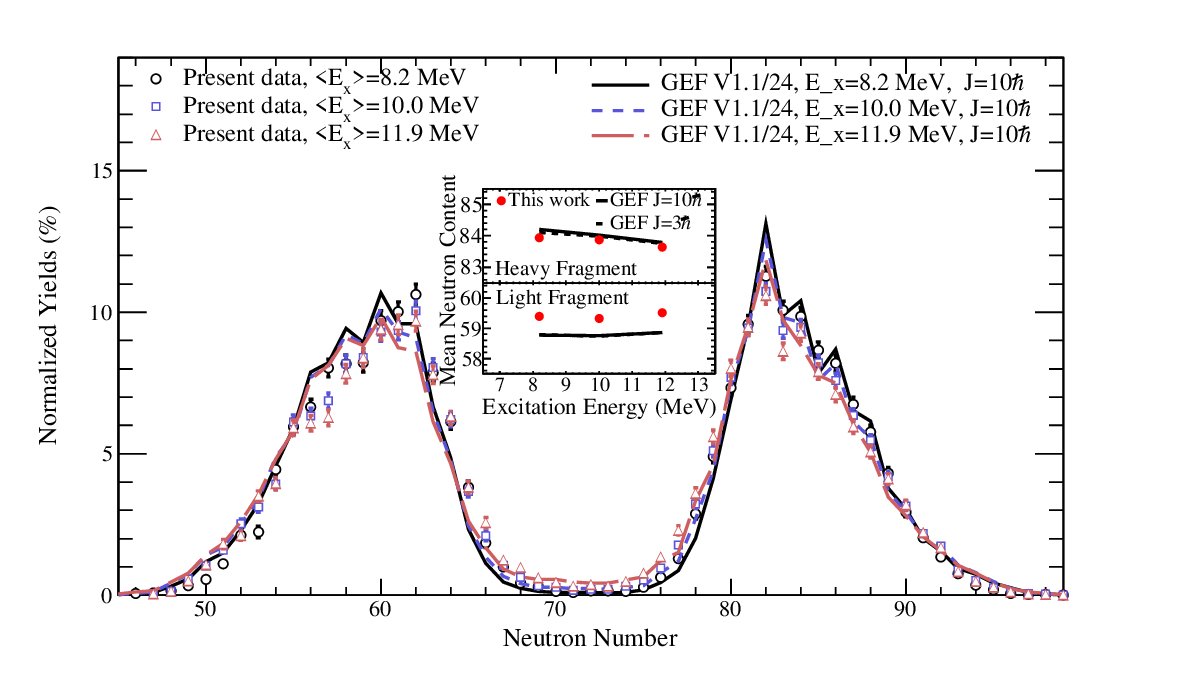}
\caption{Isotonic fission yields. Present data at three $E_x$, 8.2, 10.0,and 11.9~MeV are presented with black circles, blue squares, and red triangles, respectively. GEF calculations at the same $E_x$ are indicated with solid lines for a initial spin of $J=10\hbar$. Insets: Evolution of the mean neutron content of the heavy (top) and light (bottom) fission fragment region as a function of $E_x$. Present data (red dots) is compared with GEF predictions with initial spin $J=10\hbar$ (solid line) and $J=3\hbar$ (dotted line).}
\label{fig:YN}
\end{figure*}

The total prompt-neutron multiplicity is determined as a function of the proton number of the fragments using the neutron excess of complementary fragments, $Z$ and $94-Z$: 
\begin{equation}
\begin{aligned}
&\langle \nu_{tot}\rangle(Z,E_x)= N_{fis} -\\
&\left( Z\times\frac{\langle N\rangle}{Z}(Z,E_x)+[94-Z]\times \frac{\langle N\rangle}{Z}(94-Z,E_x)\right).
\end{aligned}
\end{equation}
where $N_{fis}=146.05\pm0.02$ is the effective number of neutrons of the compound nucleus. 

The latter takes into account the 5\% of contamination from $^{241}$Pu. Additional 0.1\% of uncertainty is included in the neutron multiplicity to take into account the neutron evaporation difference between $^{241}$Pu and $^{240}$Pu.

The resulting total neutron multiplicity as a function of the proton number of the heavy fragments is presented in Fig.~\ref{fig:Nevap}. Data sets at different $E_x$ are indicated with black circles ($E_x=8.2$ MeV), blue squares ($E_x=10.0$ MeV), and red triangles ($E_x=11.9$ MeV). Present data show a maximum neutron multiplicity at symmetry and a minimum at $Z=50$. This observation is interpreted as arising from two distint scission-point configurations: an elongated and well-deformed configuration at symmetry leading to a high deformation energy released through neutron evaporation, and a more compact configuration at $Z=50$ with less available energy for neutron emission. With increasing $E_x$, neutron evaporation at $Z=50$ increases more rapidly than at larger asymmetries. This suggests that the shape and/or the length of the pathway along the potential energy landscape towards the compact scission configuration are highly sensitive to the initial $E_x$.  

Present data are compared with thermal-neutron induced fission ${}^{239}Pu(n,f)$ from Ref.~\cite{TSUCHIYA01112000} corresponding to $E_x=6.5$ MeV, and shown with a solid green line. These data were converted from pre-neutron-evaporation fission-fragment masses to proton numbers using the fission-fragment neutron excess at scission calculated with GEF. Contrary to expectations, neutron-induced fission data at $E_x=6.5$~MeV exhibit higher neutron multiplicity than the present data at $E_x = 8.2$~MeV, with the difference decreasing for larger asymmetries. As discussed above, this may be driven by the additional angular momentum present in the system, induced by the two-proton-transfer reaction, compared to neutron-capture reactions.

Calculations from GEF do not fully reproduce the behavior observed in data. They are also shown in Fig.~\ref{fig:Nevap} with solid ($J=10\hbar$) and dotted lines ($J=3\hbar$) following the same color pattern as the experimental data. The strong even-odd staggering predicted by the model is not observed in the data. Furthermore, the minimum neutron evaporation at $Z=50$ is not reproduced by the calculations; however, the increase in neutron evaporation with increasing $E_x$ is observed in both the present data and GEF.   

Figure~\ref{fig:YN} presents fission yields of $^{240}$Pu as a function of the neutron content of the fragments. Present data are shown with black circles, blue squares, and red triangles, corresponding to $E_x=8.2$, 10.0, and 11.9~MeV, respectively. In the heavy-fragment region, there is a clear effect of the neutron evaporation stopping at $N=82$, which enhances the yield relative to the light fragments. Fig.~\ref{fig:YN} also includes GEF calculations at the same $E_x$ as the present data and initial spin $J=10\hbar$. The calculation shows a very good agreement for the heavy fragments, while it is systematically shifted towards less neutron-rich isotopes in the light fragment by approximately half a neutron.

The upper and lower insets of Fig.~\ref{fig:YN} show the evolution of the mean neutron content of the heavy ($N\geq72$) and light ($N<72$) fragment distributions, respectively, as a function of $E_x$. Present data (red dots) are compared with GEF calculations at $J=10\hbar$ (solid black line) and $J=3\hbar$ (dotted black line). 
The stabilization of the mean neutron content in the light-fragment distribution indicates a negligible effect of the additional excitation energy on the neutron evaporation of the light fragment. GEF reproduces this stabilization well, although it is systematically shifted by half a neutron towards less neutron-rich isotopes compared to the data. In the heavy-fragment region, a drift towards less neutron rich isotopes with increasing $E_x$ is observed in both present data and GEF. This reflects an enhanced neutron evaporation driven by the intrinsic excitation energy of the system.

\section{CONCLUSION}

This manuscript reports the measurement of the evolution of the isotopic fission yields of $^{240}$Pu as a function of the initial excitation energy of the system near the fission barrier.

The system was produced via two-proton-transfer reactions between a $^{238}$U beam and a $^{12}$C target in inverse kinematics. The fission fragments were isotopically identified using the VAMOS++ magnetic spectrometer.

The good agreement in mass yields between present data and previous measurements from neutron-induced fission demonstrates the high quality of the data that extends the available experimental information towards higher excitation energies, in the range of fast neutrons. This new data will help to constrain fission models and evaluations, with potential impact on nuclear applications.

As a result, fission yields in the symmetric valley are observed to increase with excitation energy, indicating a clear damping of shell effects. Furthermore, simultaneous measurements of mass and proton numbers of fission fragments provide access to their neutron content and total neutron evaporation. The neutron excess of fragments in the heavy region decreases with increasing excitation energy, whereas in the light region it remains largely unaffected. This observation is consistent with a constant flow of energy from the light to the heavy fragment in a superfluid regime, which prevents the light fragment from acquiring additional excitation energy, while the excess of initial energy is dissipated through neutron evaporation from the heavy fragment only.

Additional structure effects, which are hidden in elemental fission yields, become clear in the present measurement around $Z=50$, where the neutron excess reaches a maximum and the neutron evaporation reaches a minimum, both presumably driven by the lower deformation energy of neutron-rich Sn isotopes.

The total neutron evaporation increases more rapidly at $Z=50$ than at larger asymmetry suggesting that the compact scssion configuration around Sn is highly sensitive to the initial $E_x$.

When comparing these data with the semi-empirical fission model GEF, overall agreement is observed in the evolution of the fission yields with the excitation energy. Non-negligible discrepancies are, however, observed in detailed neutron content of light fission fragments, where the model predicts isotopes that are on average approximately half a neutron less neutron-rich that those measured in the present work.

The present measurement is limited by a reduced excitation-energy resolution that prevented a finer sampling of the evolution of the fission yields with the excitation energy around the fission barrier. This limitation is currently overcome with the development of the new-generation experimental setup PISTA, that improves the excitation-energy resolution by a factor of four~\cite{LucasPHD}. New measurements with higher precision, covering a large number of fissioning systems, have recently been performed and are planned in the near future using the combined VAMOS++ and PISTA setup~\cite{CoboProceeding}.


\begin{acknowledgments}
This work was partially supported by the Spanish Ministry of Research and Innovation under the budget items FPA2010- 22174-C02-01 and RYC-2012-11585. The excellent support from the GANIL staff during the experiment is acknowledged.
\end{acknowledgments}

\bibliography{Paper_Pu_PRC_RESUM} 
\bibliographystyle{apsrev4-2}

\end{document}